\documentclass[structabstract]{aa}
\usepackage{natbib}
\usepackage{lineno}
\usepackage{fixltx2e}
\usepackage[table]{xcolor}
\usepackage{multirow}
\usepackage{graphicx}
\usepackage{hyperref}
\bibpunct{(}{)}{;}{a}{}{,}
\usepackage{txfonts}

\newcommand{\fermi}{${\it Fermi}$}
\renewcommand{\figurename}{Figure}

\begin{document}

\title{Gamma-ray follow-up studies on $\eta$ Carinae}
\subtitle{}
\author{K.~Reitberger\inst{1} \and O.~Reimer\inst{1,2} \and A.~Reimer\inst{1,2} \and M.~Werner\inst{1} \and K.~Egberts\inst{1} \and H.~Takahashi\inst{3}}
\institute{Institut f\"{u}r Astro- und Teilchenphysik and Institut f\"{u}r Theoretische Physik, Leopold-Franzens-
Universit\"{a}t Innsbruck, A-6020 Innsbruck, Austria \and Kavli Institute for Particle Astrophysics and Cosmology,
Department of Physics and SLAC National Accelerator Laboratory, Stanford University, Stanford, CA
94305, USA \and Hiroshima Astrophysical Science Center, Hiroshima University, Higashi-Hiroshima, Hiroshima 739-8526,
Japan}
\date{Received ... ; accepted ...}

\abstract
{}
{Observations of high-energy $\gamma$-rays recently revealed a persistent source in spatial coincidence with the binary system $\eta$ Carinae.
Since modulation of the observed $\gamma$-ray flux on orbital time scales has not been reported so far, an unambiguous identification
was hitherto not possible. Particularly the observations made by the \textit{Fermi} Large Area Telescope (LAT) posed additional questions
regarding the actual emission scenario. Analyses show two energetically distinct components in the $\gamma$-ray spectrum, which are
best described by an exponentially cutoff power-law function (CPL) at energies below 10 GeV and a power-law (PL) component dominant at higher energies.}
{The increased exposure in conjunction with the improved instrumental response functions of the LAT now allow us to perform a more detailed investigation of location, spectral shape, and flux time history of the observed $\gamma$-ray emission.
} 
{We detect a weak but regular flux decrease over time. This can be understood and interpreted in a
colliding-wind binary scenario for orbital modulation of the $\gamma$-ray emission. We find that the spectral shape of the $\gamma$-ray signal agrees with a single emitting particle population in combination with significant absorption by $\gamma$-$\gamma$ pair production.}
{We are able to report on the first unambiguous detection of GeV $\gamma$-ray emission from a colliding-wind massive star binary. Studying the correlation of the flux decrease with the orbital separation of the binary components allows us to predict the behaviour up to the next periastron passage in 2014.}{}


 \maketitle

\section{Introduction}

Owing to its dramatic history (the `great eruption' of 1843, subsequent disappearance, envelopment in a huge bipolar nebula, etc.),
$\eta$~Carinae ($\eta$~Car) has been of interest for astronomical and astrophysical research for more than a century. Recent observations
by AGILE \citep{Agile} and \fermi-LAT \citep{TakaPaper} established high-energy $\gamma$-ray emission in spatial coincidence
with the location of $\eta$~Car. Physical association, e.g. through observation of correlated multiband variability or orbital 
modulation of the $\gamma$-ray signal, has not yet been confidently presented. If this identification can be established, the $\eta$ Car system would be the first massive star colliding-wind binary system (CWB) in the $\gamma$-ray sky, in contrast to known $\gamma$-ray binaries, which all contain a compact object.

It is widely accepted \citep[e.g.,][]{Damineli} that $\eta$~Car is a binary system consisting of two massive stars (one being a member of the rare stellar class of luminous blue variables (LBV), the other being an O or WR star) with an orbital
period of 5.54 yr. Stellar and orbital parameters as determined by various studies are listed in Table \ref{orbital}. Both massive stars in the $\eta$~Car system
are expected to produce powerful stellar winds. Mass-loss rates and terminal wind velocities (as given in the table) are thought to be sufficiently 
high to form a wind-wind collision zone of shocked, hot gas, wherein particle acceleration \citep[as illustrated by e.g.,][]{Usov,Dougherty} in general and subsequent $\gamma$-ray emission 
in particular \citep[as described by e.g.,][]{Reimer2006,Pabich} can occur. 
The conditions in the wind-wind collision zone depend on the orbital phase of the binary system. Therefore, these models predict $\gamma$-ray emission that is generally modulated on orbital time scales. In addition there may exist complex dependencies, such as spectral cutoffs in the underlying particle spectrum due to excessive energy losses, anisotropy in the Inverse Compton process, or types of particle transport.

\begin{table}
\center
\begin{tabular}{c l l}
 Parameter & Value & Reference\\
\hline \hline $d$ & 2.3$\pm$0.1 kpc & \citet{Davidson1997} \\
 $P$ & 2024$\pm$2 d & \citet{Corcoran05} \\
 $i$ & 45$^\circ$ &  \citet{Okazaki08} \\
 $e$ & 0.9 &  \citet{Smith04} \vspace{1 mm} \\
 \multirow{2}{*}{$\Phi$} & 27$^\circ$ & \citet{Okazaki08} \\
  & 0-30$^\circ$ &  \citet{Parkin} \vspace{1 mm}\\
 $a$ & 15.4 AU &  \citet{Corcoran01} \\
 $M_A$ & 90 M$_{\odot}$ &  \citet{Hillier01} \\
 $M_B$ & 30 M$_{\odot}$ &  \citet{Verner05}  \vspace{1 mm} \\
 \multirow{2}{*}{$\dot{M}_A$} & $10^{-3}$ M$_{\odot}$yr\textsuperscript{-1} &  \citet{Hillier01} \\
  & $2.5\times10^{-4}$ M$_{\odot}$yr\textsuperscript{-1} &  \cite{Pittard02} \vspace{1 mm} \\
 \multirow{2}{*}{$\dot{M}_B$} & $10^{-5}$ M$_{\odot}$yr\textsuperscript{-1} &  \citet{Pittard02} \\
  & $1.5 \times 10^{-5}$ M$_{\odot}$yr\textsuperscript{-1} &  \citet{Parkin} \vspace{1 mm}\\
 $V_{\infty,A}$ & 500 kms$^{-1}$ &  \citet{Hillier01} \\
 $V_{\infty,B}$ & 3000 kms$^{-1}$ &  \citet{Pittard02} \\
\hline
\end{tabular}
\caption{Several important stellar, orbital and stellar-wind parameters of the $\eta$~Car binary system and their values as determined by various studies. $P$ and $d$ are the orbital period and the distance from the Earth. $i$, $e$ and $a$ are inclination angle, eccentricity and semi-major axis. The parameter $\Phi$ represents the angle between the projection of the orbital plane along the line of sight and the apastron side of the semi-major axis in the prograde direction: $\Phi = 270^{\circ} - \omega$, with $\omega$ being the commonly used periastron longitude angle. $M$, $\dot{M}$ and $V_{\infty}$ are stellar mass, mass-loss rate and terminal velocity of the stellar wind for the stars A and B in the system.}
\label{orbital}
\end{table}

Previous analyses of \fermi-LAT data \citep{TakaPaper,Walter} have elaborated on the catalogued \fermi-source 1FGL~J1045.2$ - $5942 {\citep{1FGL}
that was found to be spatially consistent with the location of $\eta$~Car. Following the nomenclature of the new \fermi-LAT two-year catalogue \citep{2FGL}, the source designation 2FGL~J1045.0$-$5941 is used from here on.  The high-energy spectrum has been best described by an exponential cutoff power-law
function (CPL) that appears to be dominant below 10 GeV, and an additional component described by a power-law function (PL) at higher energies. Neither study
presented a flux time history that would unambiguously relate the observed $\gamma$-ray emission to a physical origin in the $\eta$~Car system. Whereas there is
evidence for nonthermal emission of $\eta$~Car at energies well below the MeV range \citep{Leyder,Sekiguchi}, no $\gamma$-ray emission  at energies above the energy range of the \fermi-LAT instrument has been reported from the vicinity of $\eta$~Car \citep{Hess}.

\section{Observations with the Large Area Telescope } 
\subsection{Dataset}
\label{data}
The \fermi-LAT started regular survey-mode observations on 2008 August 4, providing a complete, regular sky-coverage with approximately uniform exposure. Accordingly, the source of interest was monitored 
regularly over the whole time range of the dataset that is used in this analysis ($\sim$35 months, 2008 August 4 to 2011 July 5). The data were 
reduced and analysed using the \textsc{Fermi Science Tools v9r23} package\footnote{See the Fermi Science Support Centre (FSSC) website for details:
\url{http://fermi.gsfc.nasa.gov/ssc/}}. 
To reject atmospheric \mbox{$\gamma$-rays} from the Earth's limb, we excluded time periods in which the region was observed at a zenith angle greater than 100$^{\circ}$ and for observatory 
rocking angles greater than 52$^{\circ}$. All remaining photons with energy \mbox{E $>$ 200 MeV} within a square of 21$^{\circ}$ base length (aligned in directions of right ascension $\alpha$ and declination $\delta$) centred on the nominal location of $\eta$ Car, ($\alpha$, $\delta$)=(161.265$^\circ$, $-$59.685$^\circ$) were used.

\subsection{Likelihood analysis}
Because $\eta$~Car is located at low Galactic latitude and within the projection of the Carina-Sagittarius arm of the Milky Way, careful modelling of the diffuse $\gamma$-ray 
emission is required. Due to the size of the \fermi-LAT's point-spread function (PSF) -- the 68\% containment radius is more than 2$^{\circ}$ at 200 MeV -- possible contributions from nearby sources have to be taken 
into account as well. We performed maximum likelihood analyses using the instrument response function P7SOURCE$\_$V6 in conjunction with the Galactic diffuse model 
\emph{gal$\_$2yearp7v6$\_$v0.fits} and an isotropic background component \emph{iso$\_$p7v6source.txt}. The applied source 
model includes 38 point sources that are located within a radius of 15$^{\circ}$ around the source of interest and are listed in the  \fermi-LAT two-year (2FGL) catalogue \citep{2FGL}. For each source the same spectral model as in the catalogue was used. The spectral parameters of all sources within a radius of 10$^{\circ}$ were allowed to vary in the likelihood analysis. Most sources are modelled either by a simple PL
\begin{equation}
\frac{dN}{dE}=A\:E^{-\Gamma}
\end{equation} (where $A$ is the normalization parameter and $\Gamma$ is the spectral index)
or by a LogParabola function, which provides additional information on the spectral curvature of some sources by introducing a break energy $E_b$ and the parameters $\alpha$ and $\beta$, \begin{equation} \frac{dN}{dE}=A(\frac{E}{E_b})^{-(\alpha+\beta\log(\frac{E}{E_b}))}\: .\end{equation}  Exceptions are $\gamma$-ray pulsars, which are modelled by a CPL functional expression,
\begin{equation} \frac{dN}{dE}=A\: E^{-\Gamma}\mathrm{exp}(-\frac{E}{E_{\mathrm{cutoff}}})\: , \end{equation}
where $E_{\mathrm{cutoff}}$ is the cutoff energy (other parameters defined as above.) In all three cases $\frac{dN}{dE}$ gives the differential flux.\\

Source detection significance can be described by the likelihood test statistic value $TS=-2\ln(L_{max,0}/L_{max,1})$, which compares the ratio of two values that are obtained by a maximum likelihood procedure. $L_{max,0}$ is the likelihood for a model without an additional source at a specified location (the null-hypothesis) and $L_{max,1}$ is the maximum likelihood value for a model including an additional source. Surrounding sources and background are taken into account in both cases. The same notion can  also be applied in comparing the likelihood of two source models that are described by  
different spectral models for a specific source of interest. Because the TS-value is expected to follow a $\chi^2$-distribution for the difference in the number of degrees of freedom between the two models \citep{Mattox}, it can be converted into the detection significance $\sigma$ for a point source.

\section{Analysis details and results} 

\subsection{Spatial analysis} 

In the two Galactic coordinate grids in Figure \ref{L}, the TS-values indicate detection significances for a trial additional point source (a PL with two free parameters) considered sequentially at each of the grid positions.
The left figure covers the region around 2FGL~J1045.0$-$5941 at energies 0.2 to 10 GeV, the right figure at energies 10 to 300 GeV -- each representing 35 months of data. The energy band has been divided according to the previous indication of two different emission components. Both TS-maps were obtained with the tool \textit{gttsmap}. 
In addition, confidence contours mark the 68.3\%, 95.4\% and 99.7\% uncertainty regions for the location of the source defined by the maximum TS value in the grids.
For both energy bands the nominal position of $\eta$~Car is found within the 95.4\% error region. For the low band it is contained within the 68.3\% error region. The formal significance for the low- and high-energy band is 46 $\sigma$ and 12 $\sigma$, respectively.

In the lower-energy band the $\gamma$-ray signal is centred at ($\alpha$, $\delta$)=(161.243$^\circ$, $-$59.681$^\circ$) with a fairly circular 95\% confidence region of radius $r$=$0.023^\circ$. In the higher-energy band the signal is centred at ($\alpha$, $\delta$)=(161.209$^\circ$, $-$59.710$^\circ$) with an elliptical 95.4\% confidence region that has a semimajor axis of $a$=0.053$^\circ$ (NE-SW oriented) and a semiminor axis of $b$=0.045$^\circ$. The interplay between a vast photon number at lower energies and statistical limitations at higher energies yields a smaller confidence region for the hard band. The angular separations $d$ of the maximum likelihood locations for the low- and high bands and the nominal position of $\eta$~Car are $d_{\mathrm{high-low}}$ = 0.033$^\circ$, $d_{\mathrm{high-nominal}}$ = 0.038$^\circ$, $d_{\mathrm{low-nominal}}$ = 0.012$^\circ$. 
The centroid of the low-energy band is well inside the 95.4\% error region of the high-energy band. Although the high band's centroid lies slightly outside the low band's 99.7\% error circle, the error regions substantially overlap. The difference to the indicated catalogue position of 2FGL~J1045.0$-$5941 is understood by acknowledging the difference in exposure and the decomposition into low- and high-energy bands. We stress that all aforementioned positions sufficiently agree to allow interpretation in a single-source scenario.

Considering the possibility of observing the $\gamma$-ray emission superimposed by two sources in close spatial coincidence, we consider this to be extremely unlikely. Although the nominal distance between the centroid of the observed low- and the high-energy emission would not exclude source confusion between a low-energy source with exponential cutoff and a hard-spectrum source at high energies, the probability of finding two spatially coincident sources that exhibit a similar flux modulation pattern (see Section~\ref{TA}), characteristic only for objects in the rare class of long-period $\gamma$-ray binaries, appears to be the main argument for rejecting this hypothesis.

\subsection{Spectral analysis}
\label{SA}

The energy spectrum of 2FGL~J1045.0$-$5941 has been determined by a maximum-likelihood fitting technique using the tool \textit{gtlike} (dataset and source model as described above). The energy bins were chosen such that a minimum detection significance of 10~$\sigma$ per bin is required (logarithmic scale with three bins between 0.2 GeV and 1 GeV and five bins between 1 GeV and 10 GeV). Owing to statistics, the criterion was lowered to require at least 5 $\sigma$ detections per energy bin in the range from 10 GeV to 100 GeV. The highest energy bin (100 GeV to 300 GeV) corresponds to a 3 $\sigma$ detection significance. For each bin a single PL shape was assumed. The normalization parameter and the index were allowed to vary during the fitting process.
The effective energy of each bin is defined as the midpoint of the integral photon flux in the bin; therefore each energy bin carries some information about the 
underlying photon spectrum.

A spectral fit over the energy range (0.2 to 100 GeV) was obtained by fitting a source model with a CPL+PL function (both at the catalogue position of 
2FGL~J1045.0$-$5941) to the dataset up to 100 GeV using the maximum-likelihood tool \textit{gtlike}. Events at energies E $>$ 100 GeV have not been considered in fitting spectral models, because they do not significantly influence the results. Figure \ref{GlobalSpec} shows the energy spectra of 2FGL~J1045.0$-$5941 along with the overlaid broad-band spectrum. The CPL+PL multi-component functional form has been applied previously by  \cite{TakaPaper} and continues to yield better spectral representation than a single PL or CPL function. For comparison to a signal without 
a second emission component, the spectrum of the nearby pulsar PSR~J1048$-$5832 (1.2$^{\circ}$ angular distance) is shown, too. 
The best-fit parameters are a spectral index $\Gamma$=1.97 $\pm$ 0.05 and energy cutoff E\textsubscript{cutoff}=3.5 $\pm$ 0.1 GeV with an integrated flux of F$^{>\mathrm{200MeV}}_{<\mathrm{100GeV}}$ =(1.20 $\pm$ 0.04) $\times$ 10\textsuperscript{-7} cm\textsuperscript{-2}s\textsuperscript{-1} 
for the CPL, which dominates the low-energy band and $\Gamma$=1.94 $\pm$ 0.04 and an integrated flux of F$^{>\mathrm{200MeV}}_{<\mathrm{100GeV}}$ =(3.2 $\pm$ 0.3) $\times$ 10 \textsuperscript{-8} cm\textsuperscript{-2}s\textsuperscript{-1} for the PL dominant in the high-energy band. The corresponding formal significances of the two spectral components are 30~$\sigma$ and 16~$\sigma$, respectively.

\subsection{Temporal analysis}
\label{TA}
We now investigate the flux time history of the observed $\gamma$-ray emission. To search for coarse temporal signatures, two adjacent energy intervals 
were analysed: 0.2 to 10 GeV and 10 to 300 GeV, respectively. This choice corresponds to the apparent structure in the spectral energy distribution (SED) at 10 GeV as evident in Figure \ref{GlobalSpec}, 
which marks the transition between two spectral components. On the basis of 35 months of \fermi-LAT data it is not useful to search the power spectrum for the existence of 
the characteristic $\sim$5.5 yr orbital period in the $\eta$~Car system. Therefore we restrict the temporal analysis to flux variability investigations, conducted over the 
whole energy interval and in the two energy regimes separately.

\subsubsection{Flux studies in the lower-energy band (0.2 to 10 GeV)}

In Figure \ref{A}, left, we show the flux time history of 2FGL~J1045.0$-$5941 for the energy band 0.2 to 10 GeV as obtained by likelihood analysis. The dataset was divided into fourteen consecutive 
intervals, each representing 2.5 months. For each time bin a CPL function (all parameters free) was fitted to the data. The average photon flux is (1.29$\pm$0.01) $\times$ 10\textsuperscript{-7} cm\textsuperscript{-2}s\textsuperscript{-1}. Taking this value as the hypothesis for a non-varying source, a $\chi^2$-test gives a probability of 99.986\% (corresponding to 3.8~$\sigma$) that this hypothesis is false. Conducting the same study on the two nearby pulsars PSR~J1048$-$5832 and PSR~J1044$-$5737 (at angular separation of 1.2$^{\circ}$ and 2.0$^{\circ}$), which are expected to show a steady flux behaviour, we obtain merely a 1.3 $\sigma$ and 0.8 $\sigma$ significance of null-hypothesis violation. 

We conclude that in contrast to the nearby pulsars, the $\gamma$-ray source  2FGL~J1045.0$-$5941 does show a significant degree of variability in the low-energy band. Detailed analysis shows that this is mainly because of the low flux values obtained for the most recent 12.5 months of our data sample. The mean flux for this period lies about a factor 1.7 below the mean flux of the remaining data sample. The nearby pulsars do not show this downward trend.

The corresponding cumulative TS-value evaluation for 2FGL~J1045.0$-$5941 is shown in Figure \ref{A}, right. A non-variable source 
is expected to show a linear increase in the cumulative TS-value \citep{Mattox} and deviations from that can be an indication of statistical fluctuations, 
stochastic flux variability, or periodic flux variation. 
In addition to minor deviations, the figure indicates an approximately linear rise in the cumulative TS-value for the first 22.5 months of 2FGL~J1045.0$-$5941 data (blue dotted line).
Then, however, the growth in TS decreases in correspondence to the lower flux values in Figure \ref{A}, left. Thus, the conclusions motivated by the light curve are reinforced by the cumulative TS evaluation.

\subsubsection{Flux studies in the higher-energy band (10 to 300 GeV)}
\label{tempo}

Figure \ref{B} is the equivalent of Figure \ref{A} for the energy regime between 10 and 300 GeV. Due to the lower statistics at higher energies, the time binning was increased from 2.5 to 5 months. The first two bins -- representing the flux of the first ten months of 
\fermi-LAT observation - are about a factor of 2.5 higher than the following five bins. For the most recent five months only an upper limit could be obtained. A $\chi^2$-test shows that 
a null-hypothesis of observing a constant flux can be rejected with 7.4~$\sigma$ significance in favour of flux variability.

This conclusion is supported by the evaluation of the corresponding cumulative TS-values. Due to the cut-off spectra of the nearby pulsars there is no other source in the vicinity 
of 2FGL~J1045.0$-$5941 that can serve for comparison in this energy interval. Yet, the behaviour of the TS-value is indicative of $\gamma$-ray variability: whereas the 
cumulative TS-value constantly rises seemingly linearly and steadily over the first ten months (blue dotted line), it flattens and turns over afterwards.
A line connecting the origin of the graph and the last cumulative TS-value (green dashed line) would indicate steadily increasing detection significance over time 
and thus absence of flux variability. Because this is not the case, we have to conclude that the cumulative TS-curve 
is indicative of flux variability. However, this indication builds up only after the first ten months of data; after an initial period of constant flux, 
the $\gamma$-ray emission at energies 10 to 300 GeV begins to fade out.  

\subsubsection{Spectral evolution}

Figure \ref{C} shows two SEDs (determined as in Section \ref{SA}). The SED on the left represents the first ten months of our dataset for which an increased flux was observed 
at energies 10 to 300 GeV. The SED on the right covers the remaining 25 months.  As suggested by Figure \ref{B}, the data points of the hard component appear to be 
systematically lower in the latter part of the data set. Similarly as in Section \ref{SA}, a CPL+PL was fitted to the data for both time intervals using energies below 100 GeV. The most substantial change is to be seen in the high-energy PL component in which the index drops from $\Gamma$=1.41 $\pm$ 0.08 for the first ten months to $\Gamma$=2.06 $\pm$ 0.06 for the latter 25 months. Using the obtained spectral parameters to derive the integrated flux from 10 to 100 GeV (the contribution of both spectral functions are taken into account), we obtain F$^{>\mathrm{10 GeV}}_{<\mathrm{100 GeV}}=$ (1.2$^{+0.3}_{-0.2}$) $\times$ 10\textsuperscript{-9} cm\textsuperscript{-2}s\textsuperscript{-1} and F$^{>\mathrm{10 GeV}}_{<\mathrm{100 GeV}}=$ (0.49 $\pm$ 0.05) $\times$ 10\textsuperscript{-9} cm\textsuperscript{-2}s\textsuperscript{-1}, which implies a modulation by a factor of about 2.4. The difference in flux therefore agrees well with the findings of Section \ref{tempo}.

\subsubsection{Connection to the orbit of the $\eta$~Car system}
As indicated by the vertical line in Figure \ref{B}, the period of increased flux in the high-energy component (10 to 300 GeV) at the start of \fermi-LAT observations corresponds 
to the periastron passage of the $\eta$~Car binary system. To further illustrate this relation, Figure \ref{D}, left, shows the flux of 2FGL~J1045.0$-$5941 at energies from
10 to 300 GeV over the orbital phase of $\eta$~Car. The periods of increased flux clearly correspond to orbital phases close to periastron. Figure \ref{D}, right, shows flux 
in relation to the physical distance of the two stellar components (ranging from 1.5 AU at periastron to $\sim$30 AU at apastron, which was reached by October 2011). 
The intervals of increased flux correspond to a time in which the separation of the stars is less than about 12 AU, with a flux maximum (bin 1) just prior to the periastron 
passage. Otherwise the data indicate a clear decrease in flux as the binary components approach apastron. 
The same is shown in Figure \ref{E} for the low-energy component (0.2 to 10 GeV). In this case the flux decrease seems to commence later, with periods of lower flux corresponding to stellar separations of more than 25 AU.

\section{Discussion and summary}
\label{Disc}
As the analysis above has shown, we see a clear flux decrease in relation to the orbital state of $\eta$~Car. The decrease is significant at low- (0.2 to 10 GeV) as well as high- (10 to 300 GeV) energies, albeit showing a different time dependence (see above).
In general, the observed flux decrease can be qualitatively understood with predictions in the framework of colliding-wind binary models for high-energy \mbox{$\gamma$-ray} emission \citep{Reimer2006, Pabich}. As the stars move away from each other, the matter density -- as well as the radiation density -- decreases in the wind collision region where particle acceleration and subsequent \mbox{$\gamma$-ray} emission are thought to occur.  Although the observed flux variation does not yet allow an unambiguous association, we consider in the following that the detected emission indeed arises from $\eta$~Car for reasons explained above.

Note that source modulations of the two spectral components at the same amplitude can be fully recovered for the two energy intervals \textit{only} if the model used for the diffuse gamma-ray emission and for nearby point sources is perfect (e.g., without statistical and systematic uncertainties). Any realistic assessment of the backgrounds (e.g., for statistical fluctuation or, even more, systematic deviations between model and reality) will have an impact on the principal ability to recover source modulation. This will matter comparably less at higher energies due to the better signal-to-noise ratio. Hence, our inability to produce a perfect diffuse emission model principally limits our capability to recover variability in the low-energy component of $\eta$~Car.

The presence of phase-locked flux changes in any emission component refutes the idea of \citet{Ohm}, who suggested that a $\gamma$-ray signal from $\eta$~Car 
can be related to particle acceleration in the expanding blast wave that originated in the great eruption of 1843 when the primary star (the LBV) ejected about 10\% of its total mass in a massive outburst. Through shock interaction with the surrounding interstellar medium the blast wave that was created at that event might produce charged particles and subsequent $\gamma$-ray emission. In such an emission scenario one would not expect to see $\gamma$-ray flux variability in connection with the orbital configuration of the $\eta$~Car binary system. 

Based on the observations with the LAT of two apparently distinct emission components, a model has been proposed  \citep{Walter} to identify the low-energy component as due to inverse Compton scattering of $\eta$~Car's photospheric photon field off
a relativistic electron population extending up to $\sim$10\textsuperscript{4} MeV, and the high-energy component above $\sim$10 GeV as due to hadronic interactions of a relativistic proton component extending up to $\sim$10\textsuperscript{4} GeV with a high-density material in the shock region (enhanced by a factor $\sim$10 with respect to the unperturbed wind density). 
This requires electron and proton acceleration at maximum rate (Bohm diffusion regime) for values of the shock magnetic field in the sub-Gauss range. 
The target material and photospheric photon density at the shock location change with orbital phase as the inverse square of the distance between the shock and the LBV. Therefore, phase-locked orbital variations of both inverse Compton and hadronic components are expected unless the intensities of the putative accelerated particle populations change along the orbit such that it counteracts this trend. We stress here that {\it both components} are expected to vary significantly with the {\it same amplitude} for an isotropic target radiation field and no significant orbital variations in the accelerated electron-to-proton number densities. If anisotropy effects were considered, even larger variability amplitudes are  expected in the inverse Compton component than in the hadronic component.
Applied to the energy spectra determined for the first 10 months and the subsequent 25 months of LAT observations (see Figure \ref{C}) for which the mean stellar separation were $\sim$8 AU and $\sim$21.8 AU, respectively, the flux amplitude is expected to decrease from the first to the second time period by more than a factor 7 in {\it both spectral components}, possibly by an even greater factor for the leptonic component. 
The observed spectra, however, reveal flux variations in the high-energy component
by a factor 2.5
while the low-energy component shows a rapid spectral decline at a few GeV with no indication for a phase dependence of
the spectral shape, spectral index, or cutoff energy.

We note that the observed low variability of the low-energy $\gamma$-ray component with phase-locked variations of the high-energy component can be plausibly interpreted as inverse Compton scattering in the photospheric radiation field of $\eta$~Car and hadronic interactions with the wind material, respectively, if the accelerated electron number density is higher around apastron than periastron
while at the same time the accelerated proton number density does not exhibit significant
orbital variations.

Motivated by the peculiar shape of the measured spectra and the complex strong radiation fields in the vicinity of $\eta$~Car, we here consider yet another scenario.
Photon-photon absorption has long been suspected to play a significant role in the dense radiation environments of massive stars. 
For instance, first indications of \mbox{$\gamma$-ray} absorption modulating a TeV spectrum have been reported from the \mbox{$\gamma$-ray} binary LS 5039 \citep{HessLS}.

For $\eta$~Car, the observed \mbox{$\gamma$-ray} spectra can be well represented by a primary differential photon spectrum $\propto~E^{-2}exp(-E/E_{cut})$ with $E_{cut}\sim 250-500$~GeV
that suffers from \mbox{$\gamma$-ray} absorption in a black-body photon field \citep{Gould} of effective temperature $\sim 5\cdot 10^5$ K (corresponding to target photon energies of $\sim 116$ eV at the maximum of the black-body radiation field) during the first ten months of observations, and decreases to $\sim 2\cdot 10^5$ K ($\sim 46$ eV) during the subsequent 25 months (see Figure \ref{abso}). The required maximum optical depths are $\tau_{\gamma\gamma,max}\approx 1.1$ and $\tau_{\gamma\gamma,max}\approx 2$, corresponding to fluxes (at the distance of $\eta$~Car) of $\sim 6\cdot 10^{-8}$~erg cm$^{-2}$s$^{-1}$ and $\sim 9\cdot 10^{-11}$~erg cm$^{-2}$s$^{-1}$, respectively. Considering the total spectrum (35 months), the required black-body absorber has an effective temperature of $\sim 3.1\cdot 10^5$ K ($\sim 72$~eV) 
and $\tau_{\gamma\gamma,max}\approx 1.3$ corresponding to a flux of $\sim 2\cdot 10^{-9}$~erg cm$^{-2}$s$^{-1}$. In this scenario, a possible origin of the absorber could be hot X-ray gas, as observed recently by XMM-Newton and Chandra \citep[e.g.,][]{XMM,Townsley}. Both experiments revealed spatially extended, structured X-ray emission components that surround the binary system. Figure \ref{geometry} (left) illustrates the geometry of such an external absorber scenario. The change of the LAT spectrum would indicate a slightly hotter rarefied plasma towards periastron than towards apastron.

In a second scenario the absorber could be linked to $\eta$~Car's wind collision (e.g., hot shocked gas). In this case (see Figure~\ref{geometry} (right))
the angular dependence of the absorption coefficient and threshold energy on the photons' collision angle has to be taken into account. The resulting optical depth therefore depends on the orientation of the wind collision region, the line of sight, and the stars of the $\eta$~Car binary system. For example, for a mean angular phase for the first 10 and subsequent 25 months of 0.973 and 0.422, respectively, and an inclination angle of $45^\circ$ and $\omega=285^\circ$, the propagation angle 
(between the collision region and line of sight) for target photons impinging from the direction of $\eta$~Car is $\sim 73^\circ$ and $\sim 119^\circ$, respectively. 
In this case we find satisfactory spectral representations for an absorber of temperature $T\approx 10^6$ K and  $\tau_{\gamma\gamma,max}\approx 1$ for the first 10 (see Figure \ref{abso}, left), and $T\approx 2.5\cdot 10^5$~K and $\tau_{\gamma\gamma,max}\approx 2$ for the latter 25 months (see Figure \ref{abso}, right).

Orbitally modulated \mbox{$\gamma$-ray} emission up to TeV energies has been observed from a number of high-mass X-ray binary systems \citep[see e.g.,][]{Hill} with orbital periods ranging from 3.4 years (PSR B1259$-$63) down to a few weeks or days (e.g., LSI +61 303, LS 5039). For the latter two systems, peak fluxes at MeV/GeV energies occur close to superior conjunction where the companion is behind the star (which also turns out to be close to periastron passage), while the flux decreases towards inferior conjunction/apastron. This behaviour is similar to the observed flux modulation in the spectrum of $\eta$~Car despite the lack of detection at VHE $\gamma$\ rays.

Ultimately, if one concludes that the observed decrease in \mbox{$\gamma$-ray} flux indeed relates to orbital conditions in the $\eta$ Carinae system, a prediction can be made for 
how the $\gamma$-ray flux will evolve until periastron is reached again. It is expected to remain at a low level until the end of 2013, when the flux should increase again and, ultimately, reach maximum shortly before the next periastron passage in mid-2014. 
The observation of this regular variability pattern will be a decisive test for current models of $\gamma$-ray emission in colliding-wind binary systems.
Until then, we anticipate that numerical simulations will consider the multitude of details not  yet taken into account in present $\gamma$-ray emission models (e.g., 
complexity in stellar radiation fields specific for the $\eta$ Car system and the Homunculus nebula). This will allow a quantitative comparison of model predictions to the observed $\gamma$-ray data over a full orbit.

\section{Acknowledgements}
\begin{acknowledgements}
The \textit{Fermi} LAT Collaboration acknowledges generous ongoing support
from a number of agencies and institutes that have supported both the
development and the operation of the LAT as well as scientific data analysis.
These include the National Aeronautics and Space Administration and the
Department of Energy in the United States, the Commissariat \`a l'Energie Atomique
and the Centre National de la Recherche Scientifique / Institut National de Physique
Nucl\'eaire et de Physique des Particules in France, the Agenzia Spaziale Italiana
and the Istituto Nazionale di Fisica Nucleare in Italy, the Ministry of Education,
Culture, Sports, Science and Technology (MEXT), High Energy Accelerator Research
Organization (KEK) and Japan Aerospace Exploration Agency (JAXA) in Japan, and
the K.~A.~Wallenberg Foundation, the Swedish Research Council and the
Swedish National Space Board in Sweden.\\
Additional support for science analysis during the operations phase is gratefully
acknowledged from the Istituto Nazionale di Astrofisica in Italy and the Centre National d'\'Etudes Spatiales in France.\\
The publication is supported by the Austrian Science Fund (FWF).
\end{acknowledgements}


\bibliographystyle{aa}
\bibliography{PaperRef}

\clearpage
\onecolumn
\begin{figure}[t]
\center
\begin{minipage}{8cm}
\centering
\includegraphics[width=8cm]{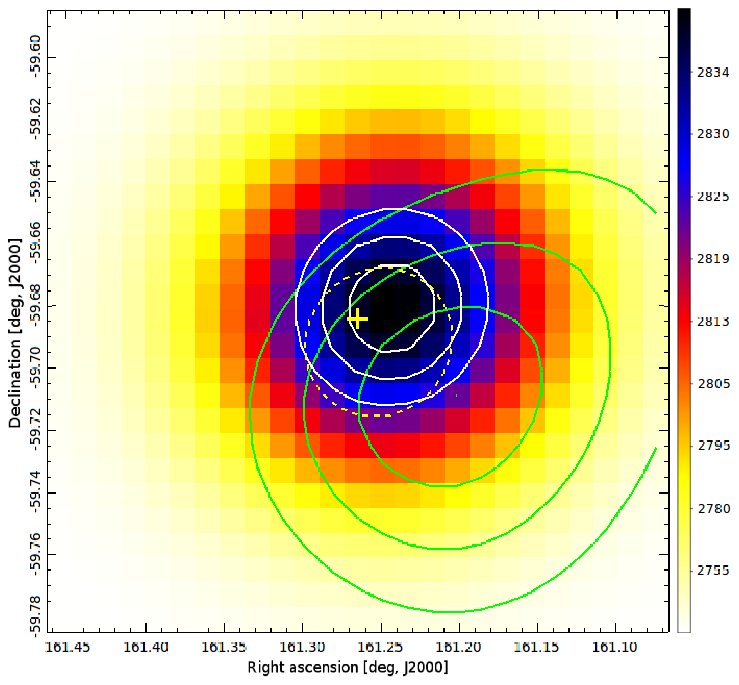}

\end{minipage}
\begin{minipage}{8cm}
\centering
\includegraphics[width=8cm]{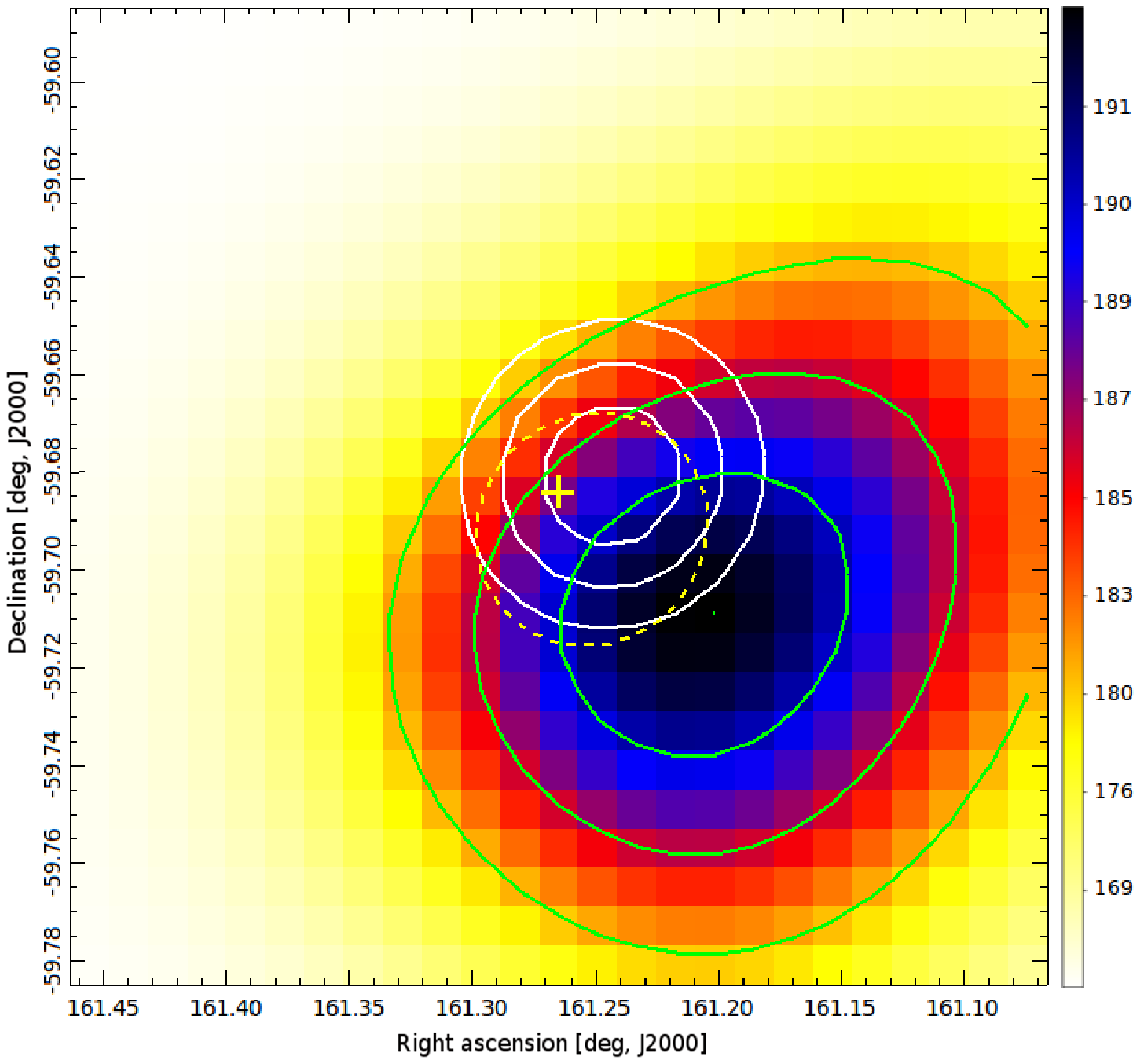}
\end{minipage}

\caption{Test statistic (TS) maps of the celestial region around 2FGL~J1045.0$-$5941 as obtained by likelihood analysis. The TS-value corresponds to the 
likelihood of the source being located in a specific grid point. Left plot: 0.2 to 10 GeV, right plot: 10 to 300 GeV. The confidence contours (green for the high band, white for the low band) mark the 68.3\%, 95.4\% and 99.7\% uncertainty regions for the location of the source of interest.
The dashed yellow circle marks the catalogued 95\% source location region of 2FGL~J1045.0$-$5941. The nominal position of $\eta$~Car is indicated by the yellow cross. \label{L} }
\end{figure}

\begin{figure}[t]

\centering
\includegraphics[width=10cm]{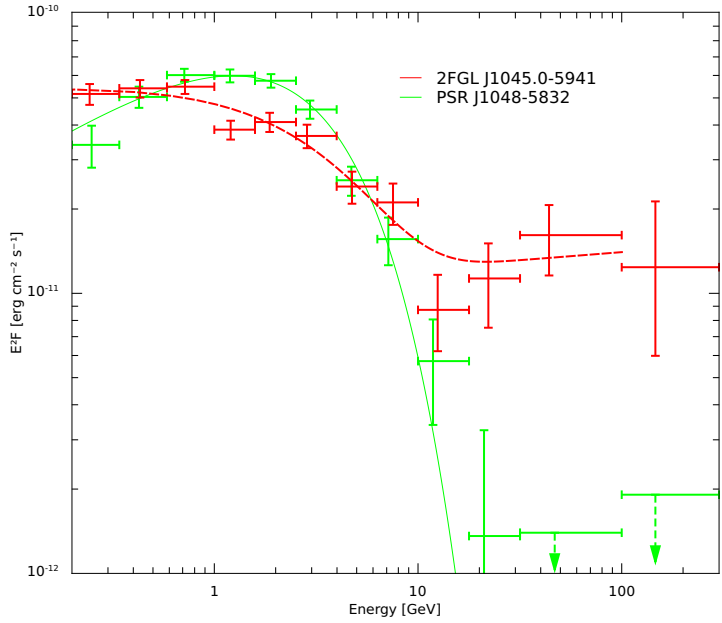}

\caption{Spectral energy distributions of 2FGL~J1045.0$-$5941 (red dashed) and PSR~J1048$-$5832 (green solid) obtained by 
likelihood analysis. The spectral model is obtained by fitting a source model with CPL and PL components at the location of 2FGL~J1045.0$-$5941 to the dataset (35 months, 0.2 $<$ E $<$ 100 GeV). The error bars are of 1 $\sigma$ type. The upper limits (represented by arrows) were determined such that the difference of the logarithmic likelihood values (with and without an additional trial point source at the indicated flux value) corresponds to 1 $\sigma$.\label{GlobalSpec}}

\end{figure}

\begin{figure}
\center
\begin{minipage}{8cm}
\centering
\includegraphics[width=8cm]{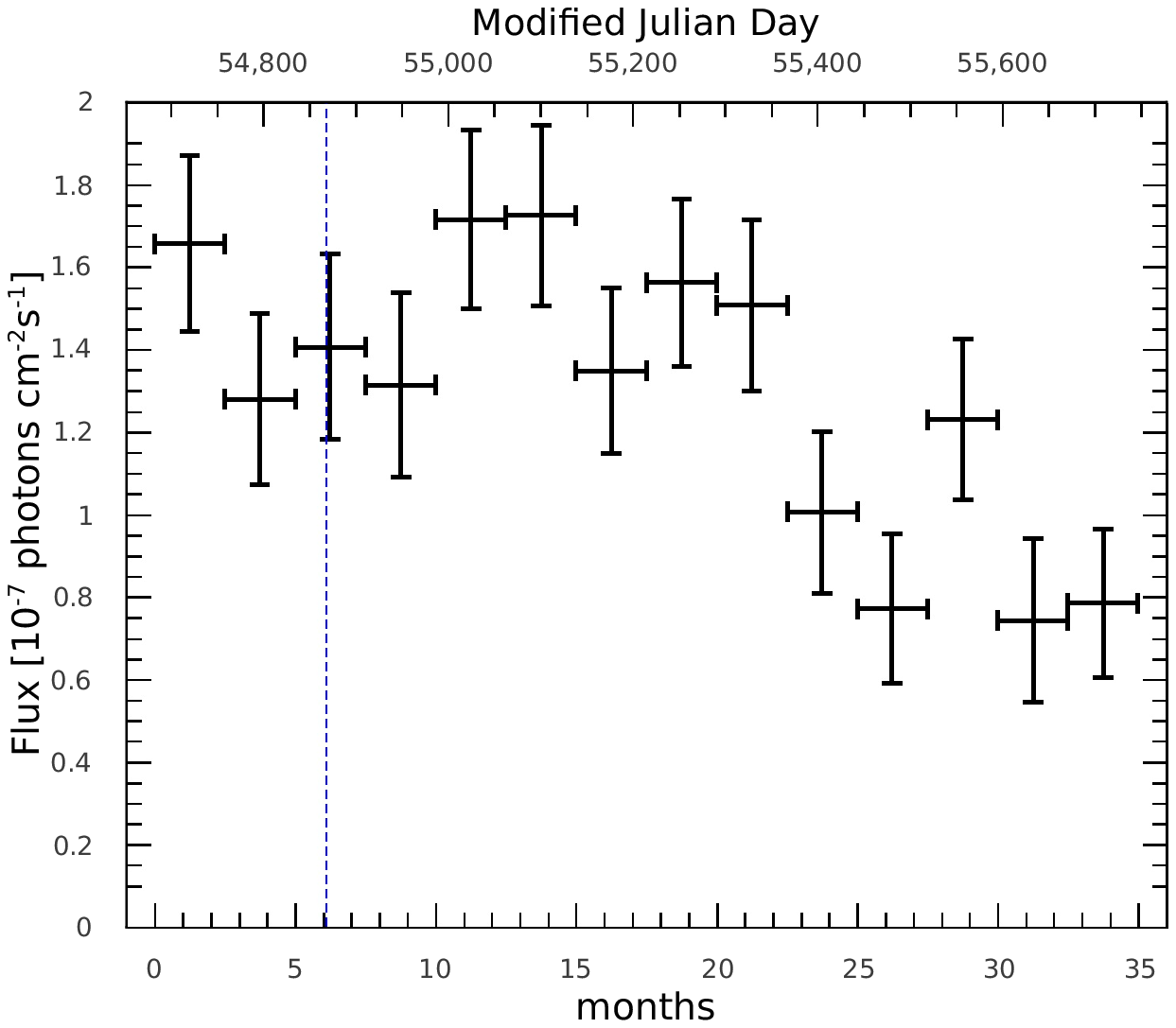}

\end{minipage}
\begin{minipage}{8cm}
\centering
\includegraphics[width=8cm]{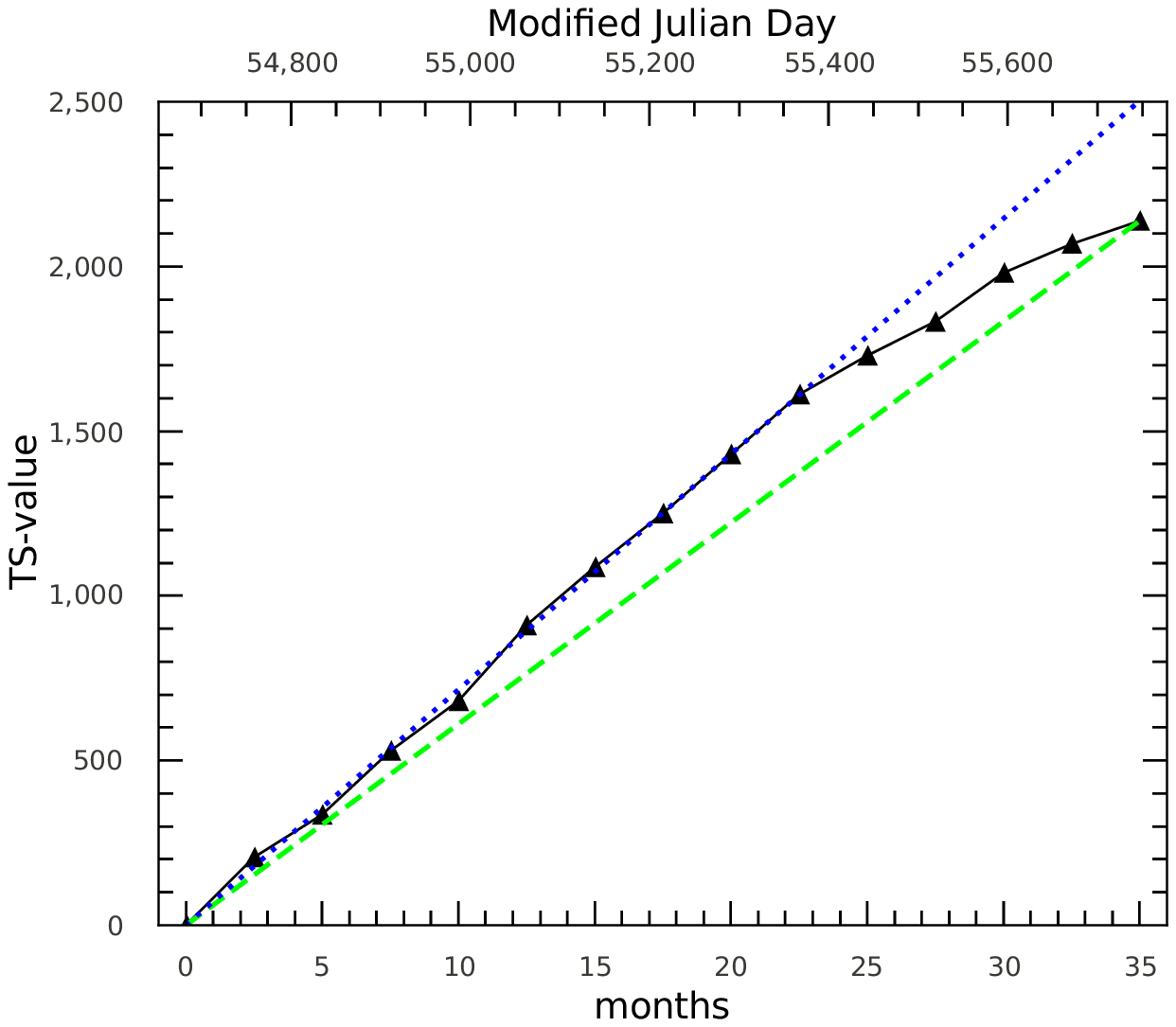}
\end{minipage}

\caption{Left: Flux time history of 2FGL~J1045.0$-$5941 in the 0.2 to 10 GeV energy band obtained by likelihood analysis. Each bin represents 2.5 months of data. 
The flux error bars are of 1 $\sigma$ type. The dashed vertical line indicates the time of the periastron passage of the $\eta$~Car binary system \citep[e.g.,][]{Parkin}. \newline
Right: Cumulative TS-value evaluation of 2FGL~J1045.0$-$5941 as obtained by likelihood analysis in the 0.2 to 10 GeV energy band. 
Each data point represents the cumulative TS-value of the source for the time interval from the start of the \fermi-LAT data taking. 
The dotted blue line is a linear fit to the first nine data points. The dashed green line connects the origin and the cumulative TS-value obtained last.\label{A} }
\end{figure}

\begin{figure}
\center
\begin{minipage}{8cm}
\centering
\includegraphics[width=8cm]{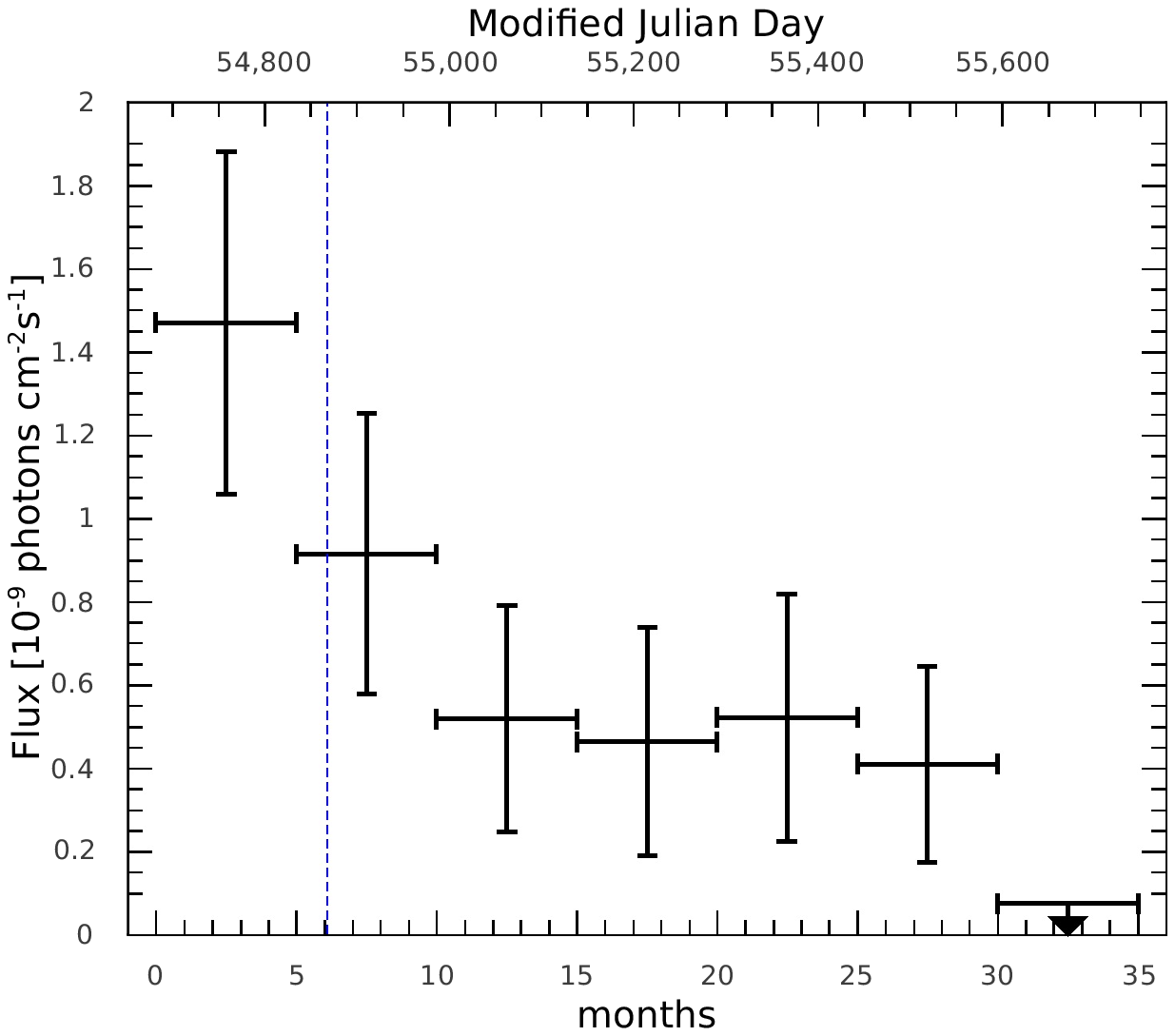}
\end{minipage}
\begin{minipage}{8cm}
\centering
\includegraphics[width=8cm]{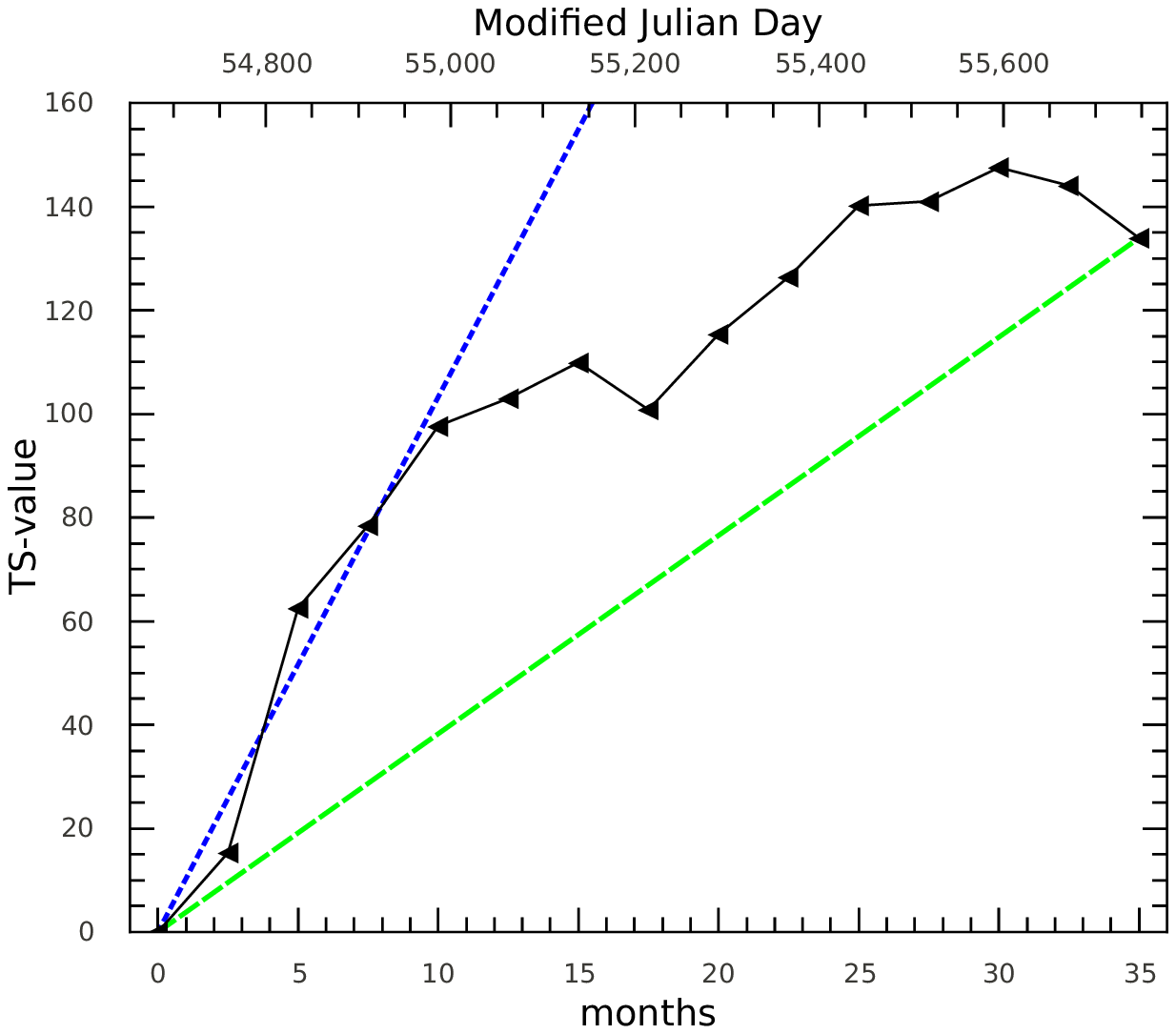}
\end{minipage}

\caption{Left: Flux time history of 2FGL~J1045.0$-$5941 obtained by likelihood analysis. Each bin represents five months of data in the 10 to 300 GeV energy band. 
The flux error bars are of 1 $\sigma$ type. The upper limit was determined as described in Figure \ref{GlobalSpec}. The dashed vertical line indicates the time of the periastron passage of the $\eta$~Car binary system. \newline
Right: Cumulative TS-value evaluation of 2FGL~J1045.0$-$5941 as obtained by likelihood analysis in the 10 to 300 GeV energy band. 
Each data point represents the cumulative TS-value of the source for the time interval from the start of the \fermi-LAT data taking. 
The dotted blue line is a linear fit to the first four data points. The dashed green line connects the origin and the cumulative TS-value obtained last. \label{B}}
\end{figure}

\begin{figure}[H]
\centering
\begin{minipage}{8cm}
\includegraphics[width=\textwidth]{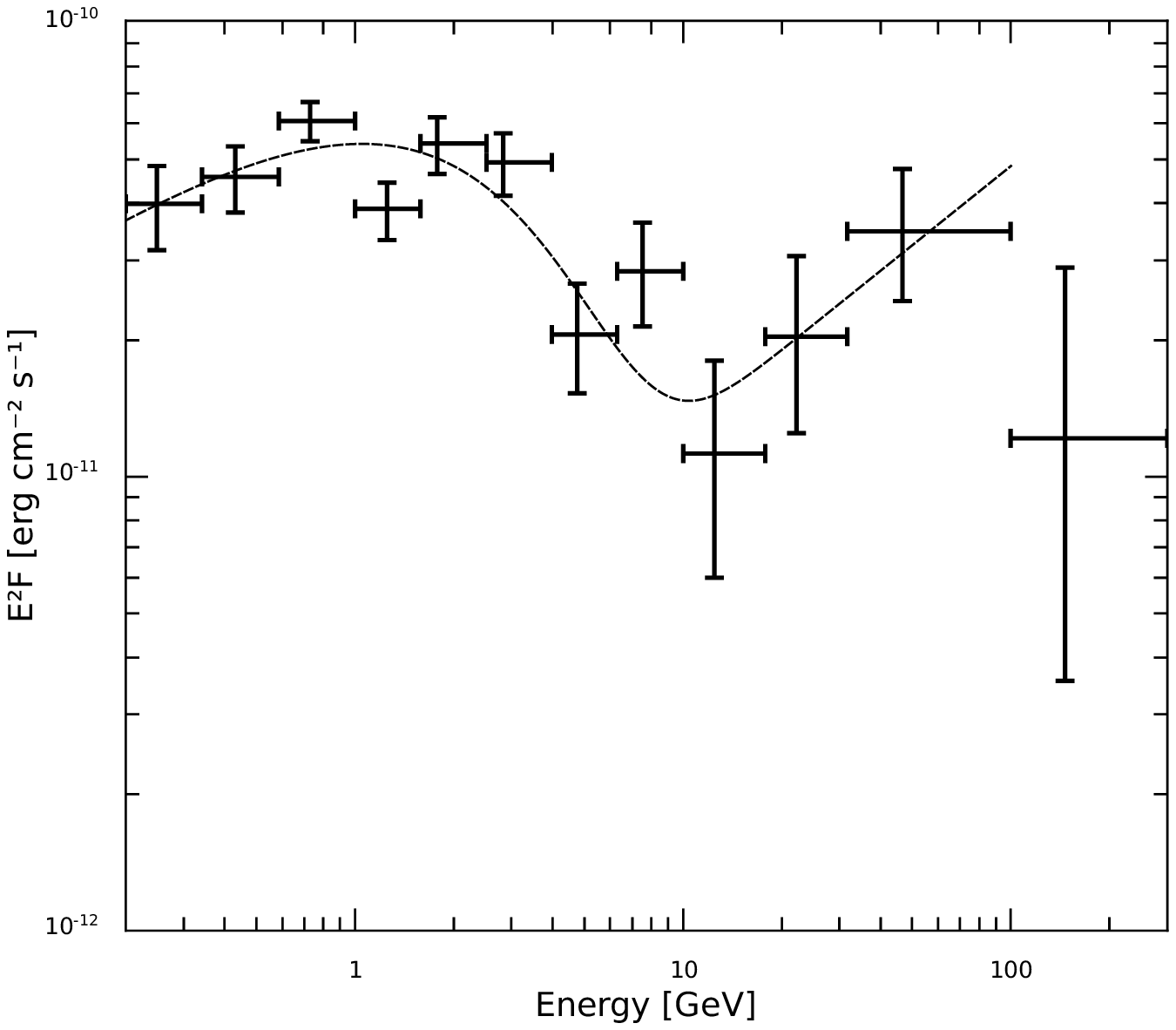}
\end{minipage}
\begin{minipage}{8cm}
\includegraphics[width=\textwidth]{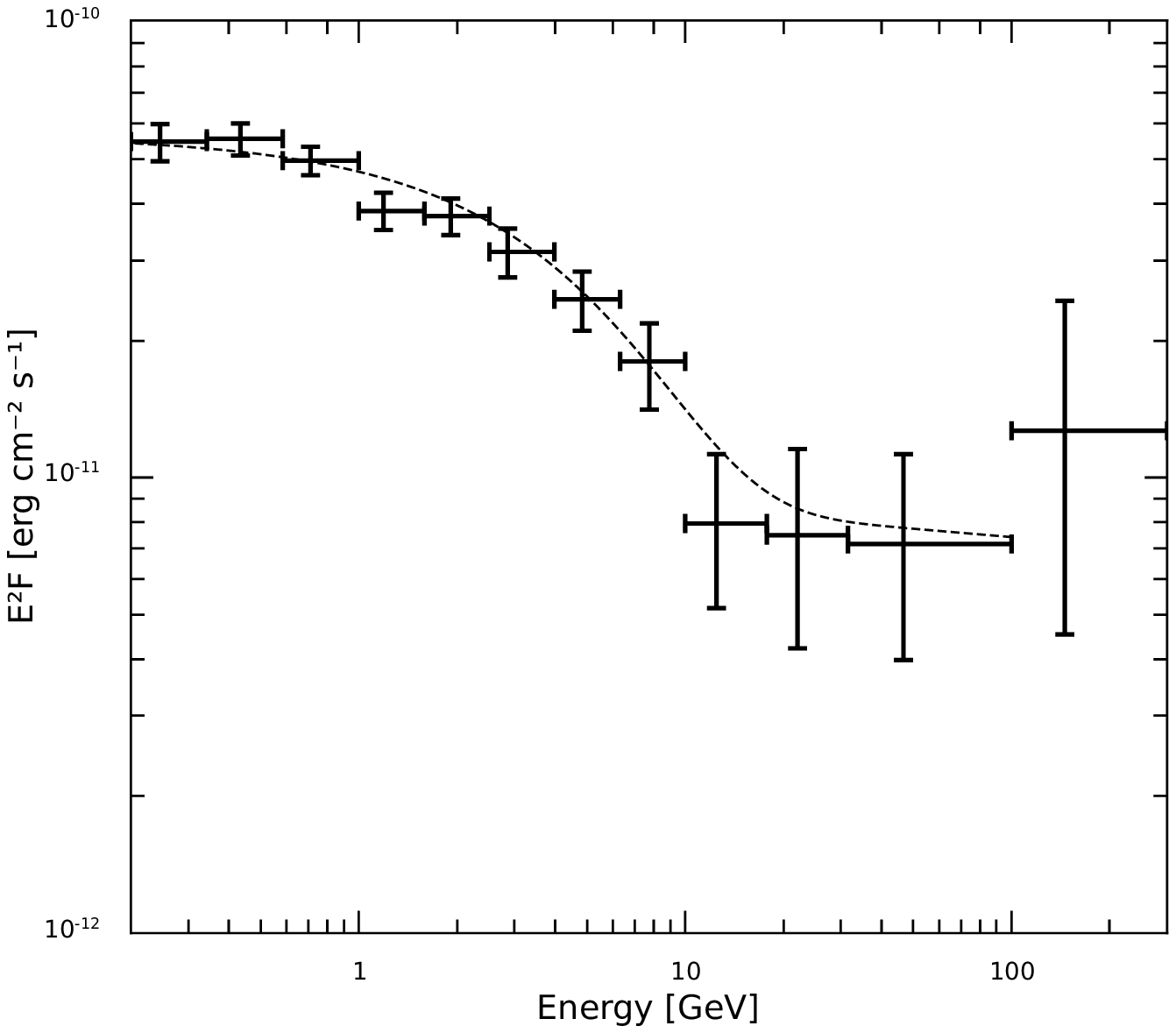}
\end{minipage}
\caption{Spectral energy distributions of 2FGL~J1045.0$-$5941 for the first 10 months (left) and latter 25 months (right) of the dataset obtained by performing likelihood analysis.
The spectral model is the sum of CPL and PL components at the position of 2FGL~J1045.0$-$5941 
to the 0.2-100 GeV data.
 \label{C}}
\end{figure}

\begin{figure}[]
\centering
\begin{minipage}{8cm}
\includegraphics[width=\textwidth]{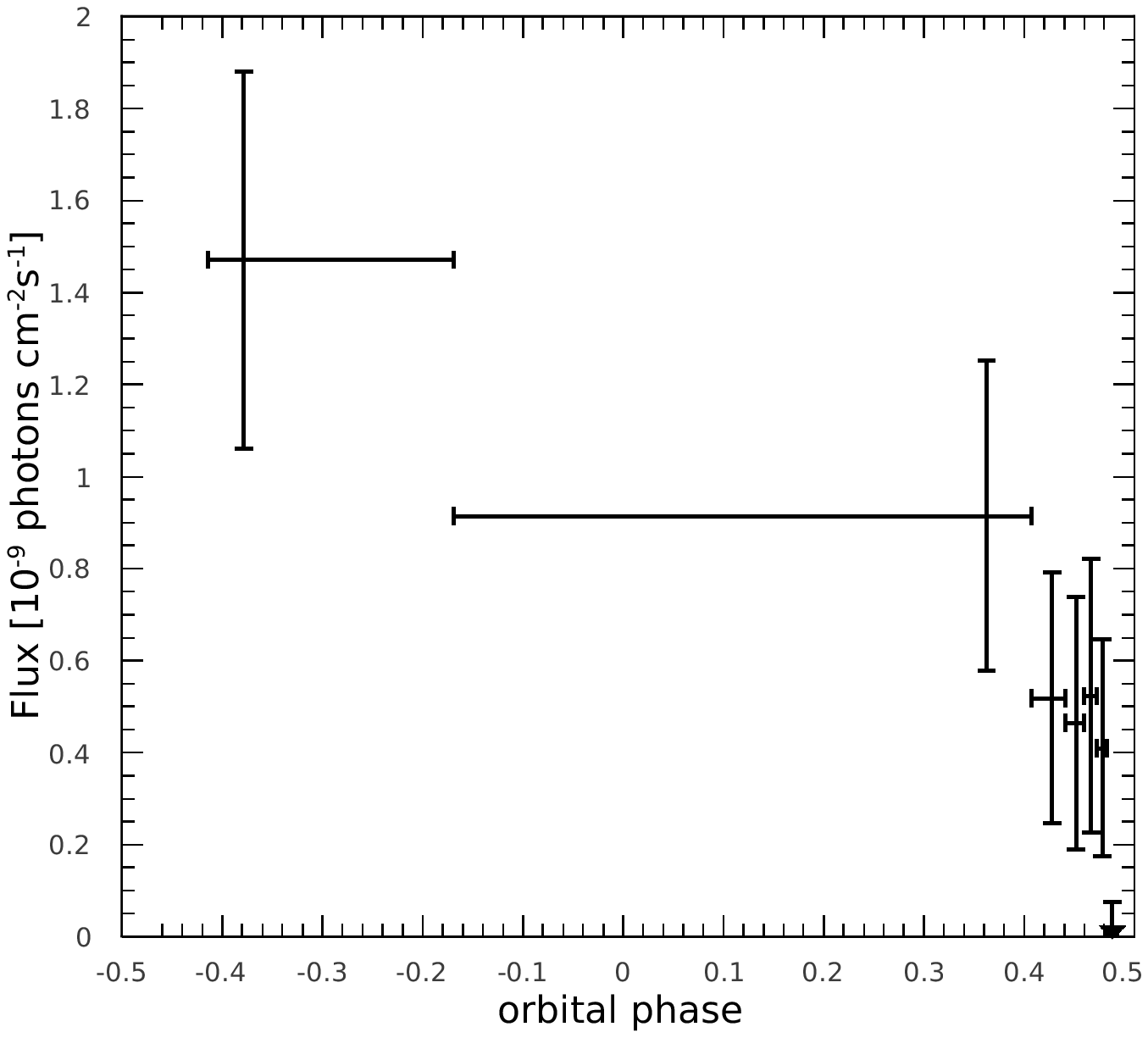}
\end{minipage}
\begin{minipage}{8cm}
\includegraphics[width=\textwidth]{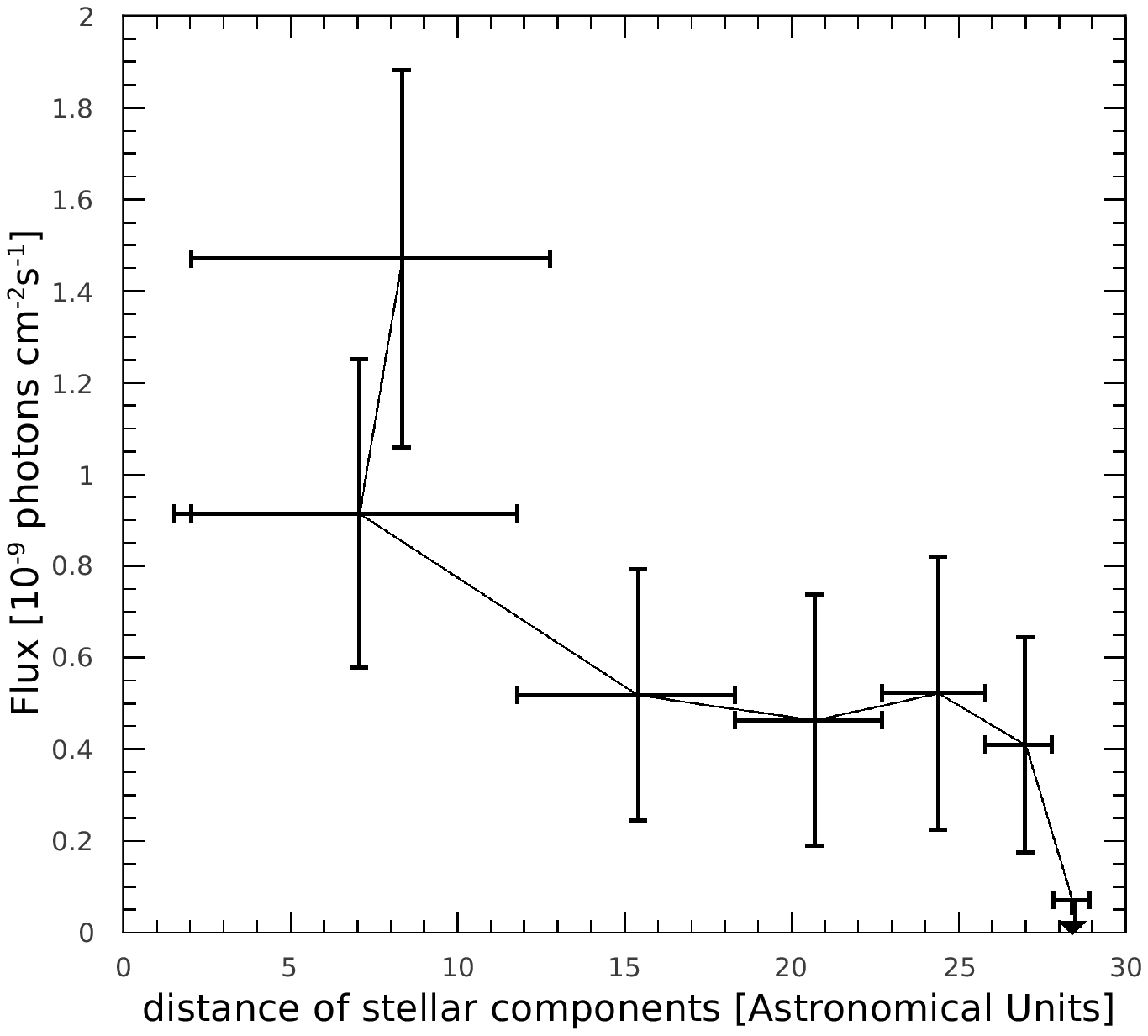}
\end{minipage}
\caption{Flux time history of 2FGL~J1045.0$-$5941 as determined by likelihood analysis for ranges of the orbital phase of $\eta$~Car (left) and the physical separation of the stellar components (right) for energies 10 to 300 GeV. Each data point represents five months of data. Because periastron is traversed very quickly, the data points around orbital phase 0 are very broad. The region between the two dashes at the left of the horizontal error bar of the second data point on the right graph is traversed twice. \label{D}}
\end{figure}

\begin{figure}[t]
\centering
\begin{minipage}{8cm}
\includegraphics[width=\textwidth]{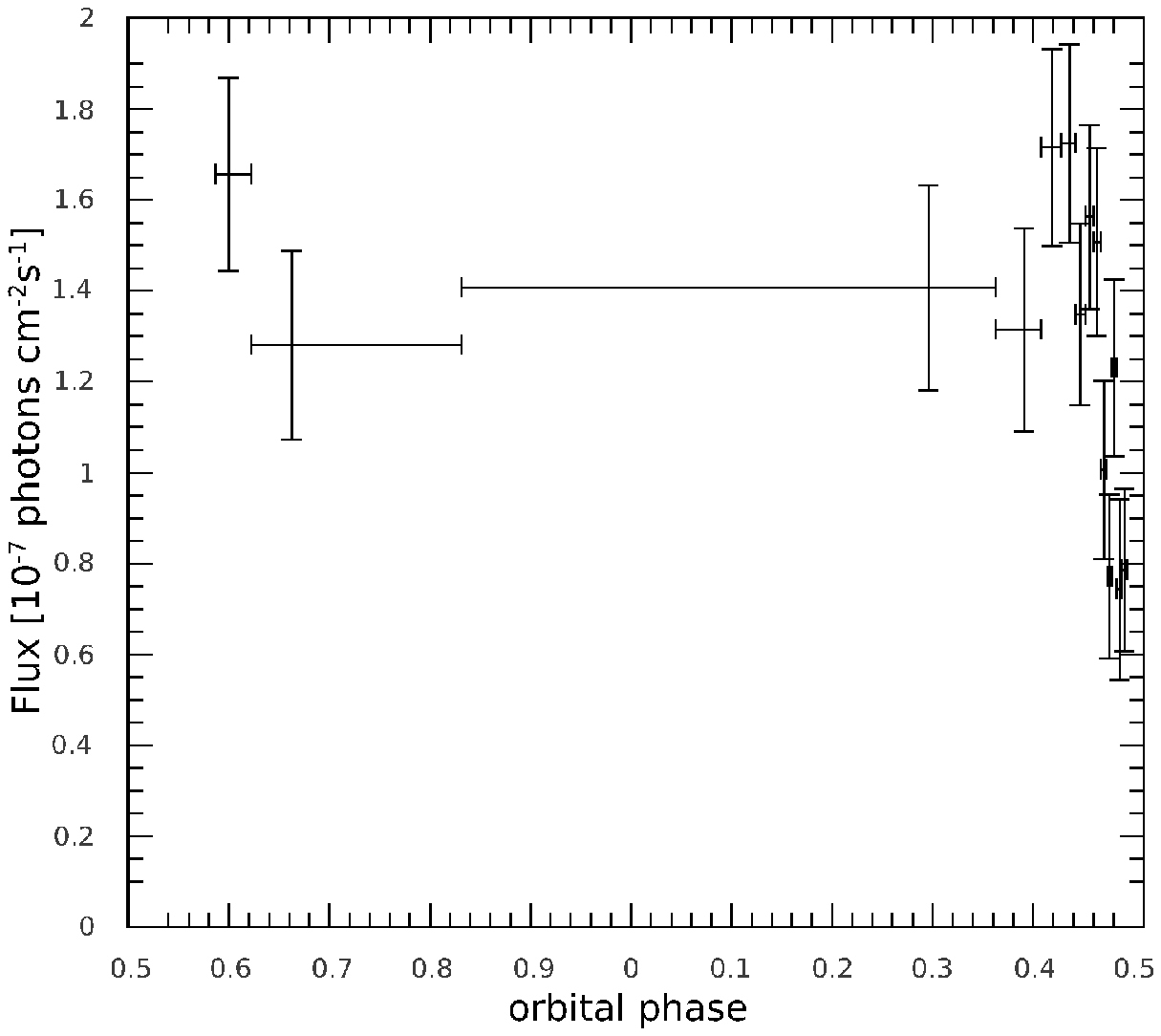}
\end{minipage}
\begin{minipage}{8cm}
\includegraphics[width=\textwidth]{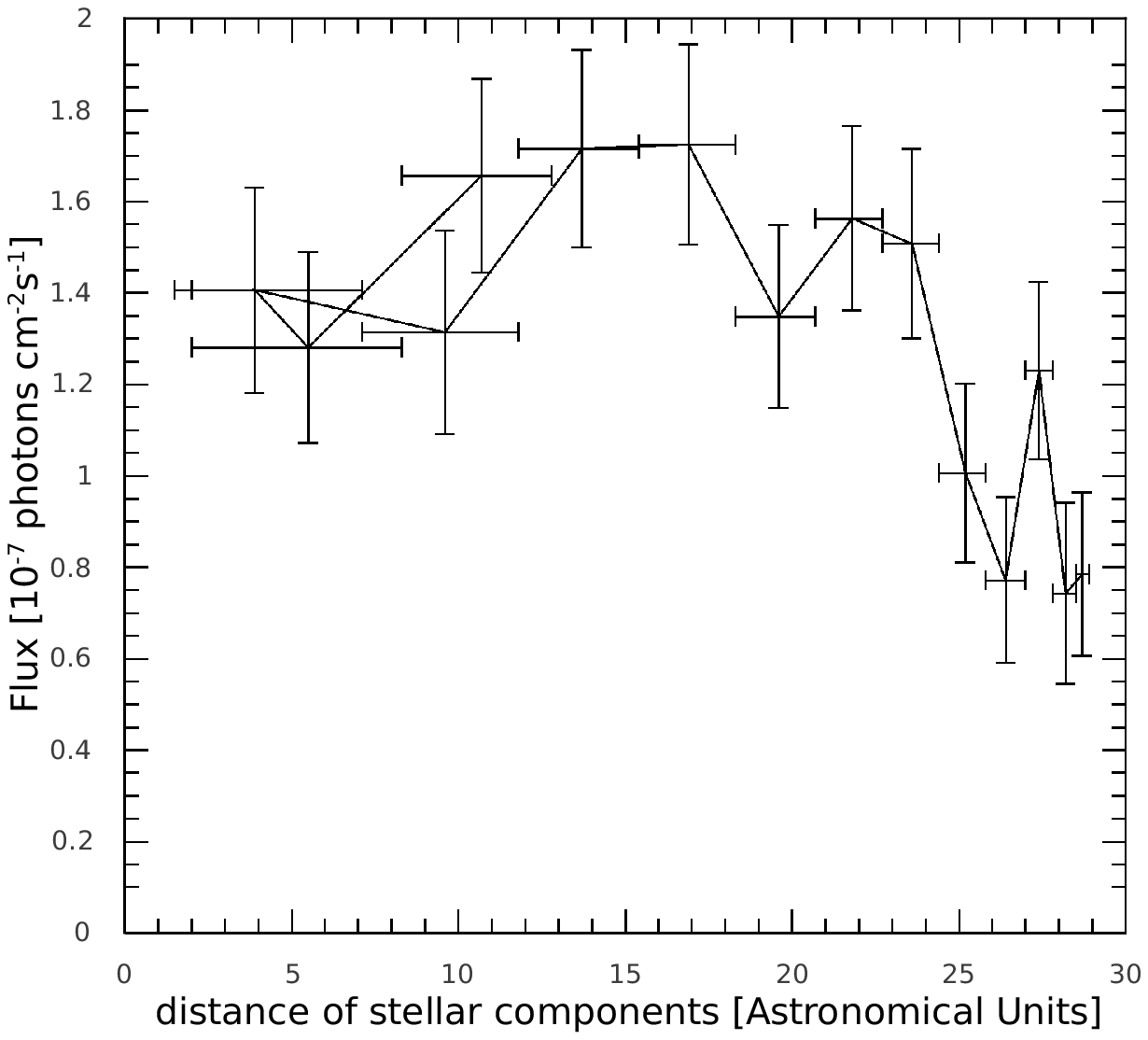}
\end{minipage}
\caption{Flux time history of 2FGL~J1045.0$-$5941 as determined by likelihood analysis for ranges of the orbital phase of $\eta$~Car (left) and the physical separation of the stellar components (right) for energies 0.2 to 10 GeV. Each data point represents 2.5 months of data. Because periastron is traversed very quickly, the data points around orbital phase 0 are very broad. The region between the two dashes at the left of the horizontal error bar of the leftmost data point on the right graph is traversed twice. \label{E}}
\end{figure}

\begin{figure}[t]
\centering
\begin{minipage}{8cm}
\includegraphics[width=\textwidth]{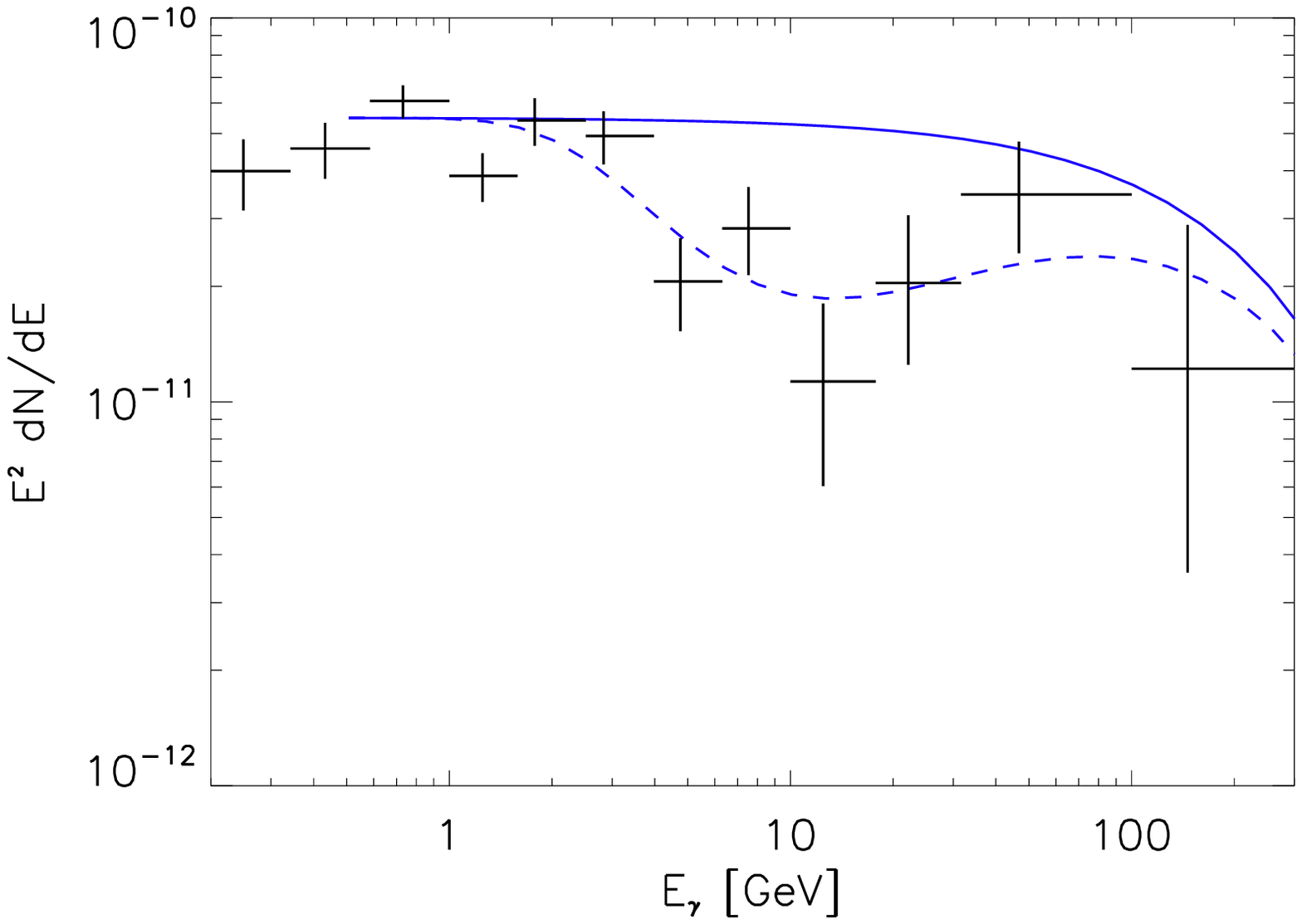}
\end{minipage}
\begin{minipage}{8cm}
\includegraphics[width=\textwidth]{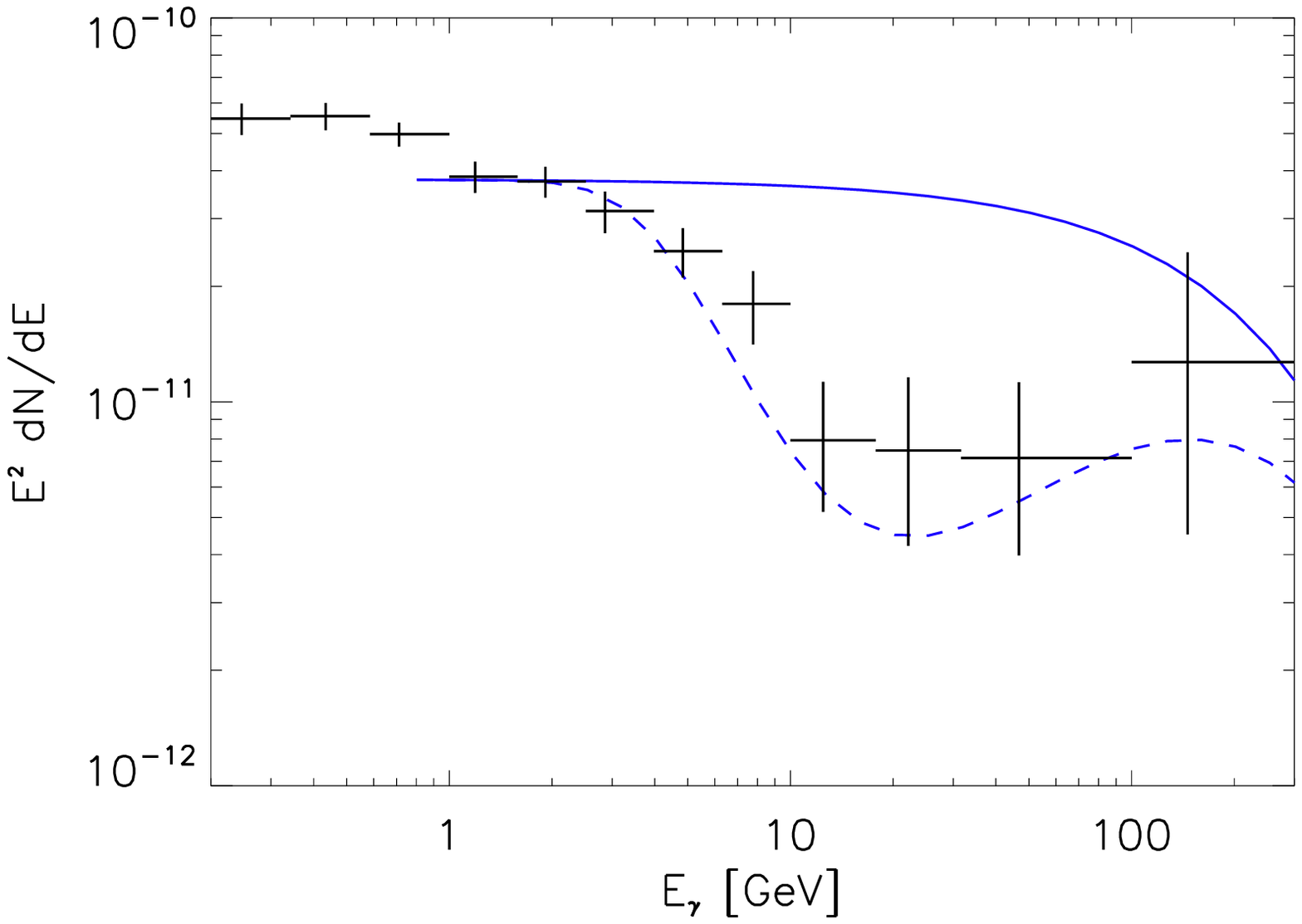}
\end{minipage}
\caption{Representation of SEDs (as in Figure \ref{C}) overlaid with the unabsorbed photon spectrum $\propto E^{-2}exp(-E/E_{cut})$ with $E_{cut}\sim 250-500$~GeV (thick line) and one that suffers from \mbox{$\gamma$-ray} absorption (dashed line) shown for the first 10 months (left) and the subsequent 25 months (right). \label{abso}}
\end{figure}

\begin{figure}[t]
\centering
\includegraphics[width=0.7\textwidth]{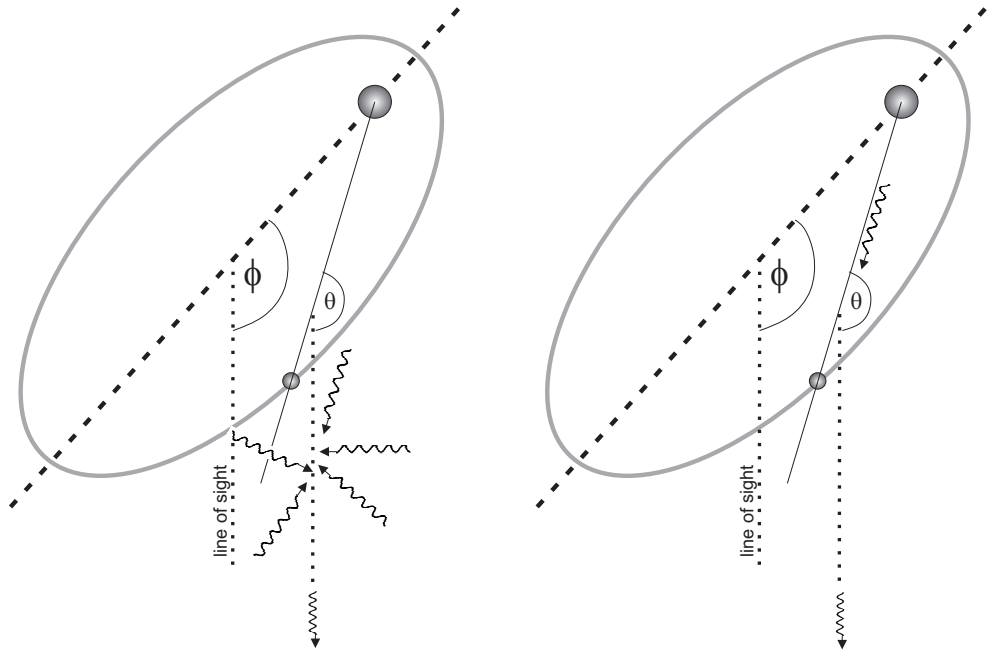}
\caption{Illustration of the projected orbit of the binary system considering $\gamma$-ray emission towards the line of sight and its possible attenuation by $\gamma$-ray absorption as described in the text. 
Scenario 1 (left): external black-body absorber (e.g., hot gas surrounding the binary system). 
Scenario 2 (right): internal absorber (e.g., hot shocked gas in the wind collision region.) \label{geometry}}
\end{figure}

\end{document}